# Bridging Nano and Micro-scale X-ray Tomography for Battery Research by Leveraging Artificial Intelligence

Jonathan Scharf[1], Mehdi Chouchane[2,3], Donal P. Finegan[4], Bingyu Lu[1], Christopher Redquest[5], Min-cheol Kim[1], Weiliang Yao[6], Alejandro A. Franco[2,3,7,8], Dan Gostovic[9], Zhao Liu[9], Mark Riccio[9], František Zelenka[9], Jean-Marie Doux[1]\*, Ying Shirley Meng[1,10]\*

[1] Department of Nano-Engineering, University of California San Diego, La Jolla, California 92093, United States of America
[2] Laboratoire de Réactivité et Chimie des Solides (LRCS), Université de Picardie Jules Verne UMR CNRS 7314, Hub de l'Energie, 80039 Amiens, France
[3] Réseau sur le Stockage Electrochimique de l'Energie (RS2E), FR CNRS 3459, Hub de l'Energie, 80039 Amiens, France
[4] National Renewable Energy Laboratory, 15013 Denver West Parkway, Golden, Colorado 80401, United States of America
[5] Department of Chemical Engineering, University of California San Diego, La Jolla, California 92093, United States of America
[6] Department of Materials Science and Engineering, University of California San Diego, La Jolla, California 92093, United States of America
[7] Alistore-ERI European Research Institute, CNRS FR 3104, Hub de l'Energie, 80039 Amiens, France
[8] Institut Universitaire de France, 75005 Paris, France
[9] Thermo Fisher Scientific, Waltham, Massachusetts 02451, United States of America
[10] Sustainable Power & Energy Center (SPEC), University of California San Diego, La Jolla, California 92093, United States of America
\* Corresponding authors: jdoux@eng.ucsd.edu (J.M.D.), shmeng@ucsd.edu (Y. S. M.)

## Abstract

X-ray Computed Tomography (X-ray CT) is a well-known non-destructive imaging technique where contrast originates from the materials' absorption coefficients. Novel battery characterization studies on increasingly challenging samples have been enabled by the rapid development of both synchrotron and laboratory-scale imaging systems as well as innovative analysis techniques. Furthermore, the recent development of laboratory nano-scale CT (NanoCT) systems has pushed the limits of battery material imaging towards voxel sizes previously achievable only using synchrotron facilities. Such systems are now able to reach spatial resolutions down to 50 nm. Given the non-destructive nature of CT, *in-situ* and *operando* studies have emerged as powerful methods to quantify morphological parameters, such as tortuosity factor, porosity, surface area, and volume expansion during battery operation or cycling. Combined with powerful Artificial Intelligence (AI)/Machine Learning (ML) analysis techniques, extracted 3D tomograms and battery-specific morphological parameters enable the development of predictive physics-based models that can provide valuable insights for battery engineering. These models can predict the impact of the electrode microstructure on cell performances or analyze the influence of material heterogeneities on electrochemical responses. In this work, we review the increasing role of X-ray CT experimentation in the battery field, discuss the incorporation of AI/ML in analysis, and provide a perspective on how the combination of multi-scale CT imaging techniques can expand the development of predictive multiscale battery behavioral models.

# 1. A Brief History of X-ray Computed Tomography

## 1.1 Introduction

X-ray Computed Tomography (X-ray CT) is well known in the medical and scientific research communities as a non-destructive imaging technique where contrast originates from the materials' absorption coefficients[1]. The attenuated X-ray beam due to the sample interaction is collected, converted, and reconstructed with sophisticated algorithms to produce cross-sectional and 3-dimensional images[2–6]. The resultant data provides valuable non-invasive information about a sample's morphology and internal structure. In the medical field, CT has led to countless discoveries and treatments that have greatly impacted the health of populations[7]. In the past two decades, the impact of CT has expanded outside the medical field to general metrology[8,9], and has considerably impacted the development of battery systems and other electrochemical devices[2,10].

With CT technology rapidly improving, commercial lab-based systems are now able to achieve similar resolutions to high brilliance synchrotron beamlines. However, with the increasing resolutions and applications of CT in electrochemical fields, more complex datasets are being explored, motivating the need for advanced analysis techniques to fully harness detailed insights about samples. This has led to the recent leveraging of Artificial Intelligence (AI) and Machine Learning (ML) to assist in the segmentation and analysis of complex datasets, or to act as a bridge between experimental data and multi-physics/multi-scale modeling[11–13]. As such, AI and ML have proved to be valuable tools to significantly reduce the time necessary to process large CT datasets while precisely labelling features of interest.

In this review, we explore the larger outlook of X-ray CT in the battery field and discuss how AI and ML can impact data analysis and computational modeling. In the first section, we discuss the key technological advancements that have made X-ray CT an advanced tool suitable for battery characterization. In the second section, we outline the virtues and limitations of CT for a variety of battery chemistries as well as the key morphological parameters that can be extracted from the experiments. The third section covers the methods of proper data analysis and filtering to extract these parameters and discusses the emerging uses of AI and ML in battery modelling. Finally, in the last section, we explore the perspective of the future of X-ray CT, in which AI and ML can be used in combination with other techniques to fully characterize battery systems and develop multi-physics and multi-scale predictive models.

## 1.2 Development of X-ray Computed Tomography

X-ray CT was first used in 1971, when Sir Godfrey Hounsfield performed the first patient brain CT scan[14–16]. Since then, CT has evolved from a technique primarily used in the medical community to a tool widely used across multiple disciplines in the scientific and engineering world[2,17]. To track the development of CT, **Figure 1** showcases the year of publication vs. the reported voxel size for works in the medical and electrochemical storage fields[18–20]. Here, a voxel is a 3D representation of a 2D pixel and corresponds to the smallest cube of information obtained from a scan. The spatial resolution is often thought of as at least two to three times the voxel size but can be larger due to blurring and imaging artifacts that impede the distinction of fine features.

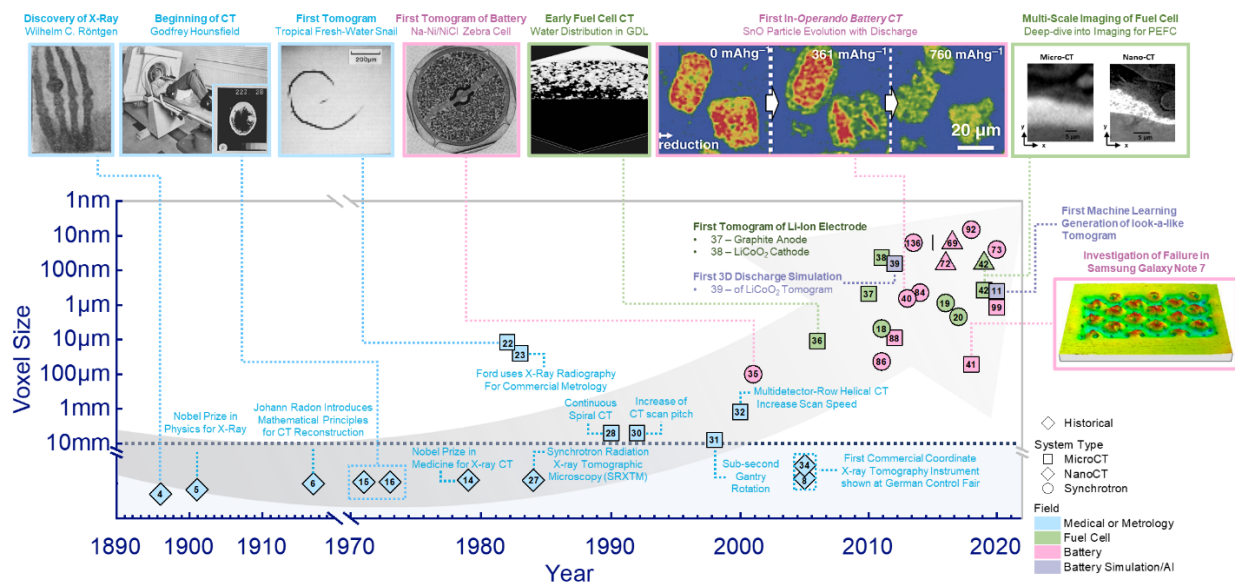

**Figure 1 | History and trends of computed tomography.** Notable advancements in X-ray CT showing the trend of decreasing tomogram voxel size with time. Colors represent the various fields works are related to: Medical (blue), Fuel Cell (Green), Battery (Pink), and Battery Simulation (Purple). Marker shapes indicate historical advancements (♦) and works pertaining to MicroCT (■), synchrotron (●), and NanoCT (▲) systems. Markers below the dotted line are notable CT-related events or works where voxel sizes are not reported.

In the early 1980s, CT started to gain traction outside the medical field in the broader scientific and industry community, and in 1982, the first *micro*-tomogram was taken of a freshwater snail with a 12 μm voxel size[21,22]. In 1983, Ford reported the first industry application of an in-house built microscale CT (MicroCT) system where they distinguished features with a spatial resolution of 25 μm[17,23]. In the same year Grodzins *et al.*[24–26] proposed the theoretical principles of utilizing *synchrotron-sourced* radiation to provide enhanced contrast and resolution for CT, and in 1984, Thomson *et al.* reported the first synchrotron radiation X-ray tomographic microscopy (SRXTM) measurement[27]. Meanwhile through the late 1980s to 1990s, developments in the biomedical community progressed laboratory-scale CT from a slow step-and-shoot approach[7] to that of continuous gantry rotation[7,28–31] with multidetector rows[32,33] that drastically decreased the acquisition speed for larger areas.

**1.3 CT Deployment in Electrochemical Device Characterization**

CT systems used in metrology benefited greatly from the improvements in speed and detection developed by medical CT. However, it wasn't until 2005 that the first CT machine dedicated to metrology and industrial applications was introduced,[8,34]; the first micro-tomograms of a battery[2,35] and fuel cell[10,36] were reported in 2001 and 2006 respectively. As shown in **Figure 1**, numerous CT works in the electrochemical field began to emerge shortly after the commercialization of this tool. In the early 2010s, the first tomograms of a Li-ion battery positive and negative electrode were reported, illustrating the capability of distinguishing the active material ($LiCoO_2$) from an inactive phase[37,38]. The first 3D discharge simulation based on tomography images quickly followed in 2012[39].

Further development allowed for novel and more creative studies to be performed: in 2013, Ebner *et al.* performed the first *operando* battery CT experiment to visualize and quantify the electrochemical and mechanical evolution of SnO particles in a Li-ion battery electrode[2,40]. In 2018, Loveridge *et al.* used X-ray CT to identify the failure mechanism in the Galaxy Note 7 which were recalled due to battery explosions[41]. The reconstructed tomograms in this work revealed defects in the positive tab welding area that resulted in electrical shorts, leading to a thermal runaway. This example illustrates the usefulness of CT and the incentive of the recent efforts toward multi-scale imaging by the electrochemical community[42]. Finally, an increasing number of works are now leveraging AI and ML, for instance for imitating *in silico* electrode tomograms similar to experimental CT[11].

**1.4 Present Day Capabilities: Lab-based and Synchrotron Sources**

For research in material science, both lab-based and synchrotron facilities have made great advancements over the past decade. In 2014, Maire and Withers wrote a review[43] on quantitative X-ray tomography where they outlined how X-ray CT data was no longer only used for qualitative insights but increasingly for quantitative analysis of material properties. This transition from qualitative to quantitative only accelerated since 2014, as imaging capabilities facilitate greater spatial and temporal resolutions. Lab-based X-ray CT systems now routinely achieve 1 µm or less resolutions, with specialized systems able to achieve resolutions as low as 10's of nm[44]. This multi-length scale capability for lab-based systems allows for *ex-situ* imaging of structural properties from 10's of nm to mm[44,45].

Recently, laboratory CT systems have enabled dual or tri-energy imaging via multiple quasi-monochromatic beam energies[46]. Software packages have been developed to extract the best detail from images taken at the different energies. For example, images taken with lower energies may have enhanced resolution and sharpness for materials consisting of low atomic mass elements. This software framework, combining images taken at multiple energies, can also be utilized for correlative workflows between different microscopy methods. The strengths of individual techniques are leveraged in combined datasets that provide heighted resolution or larger sample sizes. The temporal resolution of laboratory sources has also shown tremendous progress over the past decade but remains insufficient for many *operando* and *in-situ* analyses of structural dynamics in the range of minutes to hours. Synchrotron sources are also rapidly evolving, with major synchrotrons, such as the National Synchrotron Light Source (NSLS) II, Advanced Photon Source (APS), and the European Synchrotron Radiation Facility (ESRF) having completed or planned upgrades for increased photon flux density and coherence for faster imaging and greater sensitivity[47,48]. For example, the ESRF's Extremely Brilliant Source (EBS) is expected to present 100 times its previous brilliance and coherence, facilitating new opportunities for high energy and high spatial- and temporal resolution imaging[49]. Synchrotron sources are now achieving tomograms with voxel sizes of 20 to 50 nm in under 1 hour[50,51]. This evolution of both laboratory- and synchrotron-based capabilities has continued to present new opportunities for understanding the highly dynamic behavior of electrochemical energy devices. For consistency, in the following we refer to 3D tomographic data collected at synchrotron facilities by scanning transmission X-ray microscopy (STXM) or transmission X-ray tomography (TXM) as synchrotron radiation X-ray tomographic microscopy (SRXTM)[50,51].

## 2. X-ray CT in the Battery Field

While several tools are already routinely used to characterize the morphology of electrochemical devices or materials, X-ray CT presents significant advantages that the other techniques do not possess. For instance, Focused Ion Beam-Scanning Electron Microscopy (FIB-SEM), Transmission Electron Microscopy (TEM), and Secondary Ion Mass Spectrometry (SIMS) all require a vacuum, making *in-situ* and *operando* studies difficult if not impossible for most battery systems. Moreover, these 3D reconstruction techniques are destructive and require invasive sample preparation methods. In comparison, X-ray CT is non-destructive and does not require a vacuum for high resolution imaging, making it ideal for evaluating 3D morphological changes *in-situ* or *operando* in practical battery systems. CT can also be used to distinguish and segment species based on the varying X-ray absorption, thus allowing for select materials to be studied dynamically. A comparison of the X-ray CT with common battery characterization techniques can be seen in **Table 1**.

**Table 1 | Common tomography techniques in material science.**

| Technique | Resolution | Field of View | Destructive or Non-destructive | Vacuum Level | Information Extracted |
|---|---|---|---|---|---|
| **X-ray Computed Tomography**[52–57] | ~10 nm | ~100 µm | Non-destructive | Not Required | Porosity, Surface Area, Tortuosity |
| **Cryogenic Electron Tomography**[58] | ~1 nm | ~100 nm | Destructive | ~$10^{-8}$ kPa | 3D Nanostructures |
| **Focused Ion Beam**[59] | ~10 nm | ~10 µm | Destructive | ~$10^{-6}$ kPa | Porosity, Surface Area |
| **Atom Probe Tomography**[60] | ~1 Å | ~100 nm | Destructive | ~$10^{-11}$ kPa | Atomic Arrangements |
| **Nuclear Magnetic Resonance Imaging**[61] | ~1 mm | ~10 cm | Non-destructive | Not Required | 3D Tomography |
| **Time of Flight Secondary Ion Mass Spectrometry**[62] | ~10 nm | ~10 µm | Destructive | ~$10^{-8}$ kPa | Chemical Composition |

While X-ray CT has many benefits over other tomography techniques, the relatively limited resolution and lack of chemical information for most lab-scale CT systems make it challenging to study electrode interfaces, a crucial aspect of battery systems. Nevertheless, CT is still a relatively new tool for the electrochemical field, and as high resolution NanoCT systems are developed and used more, it will become a common battery characterization tool as high-resolution 3D imaging with precise species segmentation is now possible.

### 2.1 Materials CT Parameters

To understand the usefulness of CT in battery research, it is important to know the information and morphological parameters that it can provide. Reconstructed volumes can showcase large-scale device architecture, such as by Yin and Scharf *et al.*, who demonstrated uniform electrode contact when flexing a printed Zn-AgO battery[52]. However, apart from large-scale architecture, many morphological parameters can also be extracted, providing powerful insights into electrode structure and performance. These morphological parameters can come in a variety of types: surface area, volume, particle size distribution, or porosity and pore networks.

Surface area and volume information are commonly used in battery research to analyze electrodes and are the most intuitive to observe and quantify from reconstructed volumes. Surface area measurements can evaluate electrode wettability, while volumetric analysis can determine the thickness variation and volume expansion during battery operation. In the first *operando* battery CT study, Ebner *et al.* evaluated the volume expansion of a SnO electrode as it was lithiated[40]. During reduction, the repeated measurements using SRXTM revealed a 250% volume expansion due to lithiation that was only partially recovered during oxidation. The thickness and the spatial distribution of the solid-electrolyte interface was also quantified, showing its increase with lithiation. CT volume extraction has also proven particularly useful for silicon anode batteries[63–65] in the evaluation of volume expansion during lithiation (up to 280% for $Li_{15}Si_4$)[66], which is one of the main limitations hindering cycle lifetimes. In a 2019 study, *in-situ* SRXTM was used to track the expansion and contraction dynamics of Si electrodes during electrochemical cycling[66]. The thickness variation and changes in the delaminated area were studied, and the micro-sized crack volume fraction was quantified to reveal the failure mechanism in non-maturated electrodes.

Particle analysis can also be performed using CT: through segmentation (See **Box 2**), particles can be separated, and their individual volumes can be analyzed to provide valuable information about their size and distribution. This type of analysis can be especially useful for *in-situ* or *operando* studies, where the morphological evolution of active material particles can be tracked and analyzed dynamically. For instance, Gent *et al.* studied the heterogeneity of lithiation in secondary particles in causing accelerated capacity fade[67]. Additionally, Zernlike phase contrast (ZPC)[68] is a technique often used in NanoCT which uses phase differences in the transmitted X-ray signal to differentiate materials, and thus has been employed to segment out the three phases (*e.g.* active material, binder, and pore) typically observed in LIB electrodes[69,70]. With a high spatial resolution of 50 nm, Babu *et al.* characterized the particle contact area variation with additives to illustrate their influence on the electrode contact resistance. This can be particularly valuable in quality assurance to study the influence of manufacturing and synthesis conditions on particle morphology. For instance, Heenan *et al.* showed how 5-min long scans with a NanoCT system was sufficient to fully resolve cathode particles and directly quantify the variation in the particle's asymmetry, sphericity, and local surface roughness[71]. This study also quantified the internal voids within individual particles which should be minimized to maximize volumetric energy density.

Similarly, pores and void spaces in battery electrodes can be visualized and quantified with CT. For instance, Frisco *et al.* quantitatively extracted the pore distributions in commercial Li-Ion cells, revealing a collapse of the anode pore structure during cycling in the first investigation of SEI build up using NanoCT[72]. They showed a decrease of more than half the pore volume with cycling, and qualitatively demonstrated, using 3D tomograms, the SEI build up resulting in increased cell impedance[72]. Similarly, Su *et al.* used *operando* SRXTM to perform the first characterization of Li-$O_2$ battery cathodes with 3D tomography and extracted the pore distribution using an interconnected pore model for the scanned $Li_2O_2$ electrode[73]. By using ZPC, they were able to image and distinguish the void spaces from the lighter species like carbon and $Li_2O_2$ discharge products, successfully extracting nano-sized pores on the order of 100 nm.

Once segmented, an interconnected pore network model (PNM) can be extracted and utilized to the study the mesostructure evolution during fabrication processes such as calendering[74]. For instance, Torayev et al. introduced a 3D-resolved PNM extracted from a CT image of a Li-$O_2$ battery carbon electrode[75]. The extracted pore network consists of a family of spheres with different sizes connected by cylindrical throats and describes species transport through the electrode. Thanks to models such as this, researchers have shown that electrode samples that have the same average porosity and tortuosity factor but different pore interconnections, can result in differing discharge performances.[76,77].

Finally, CT can also allow to measure the tortuosity or tortuosity factor (square of tortuosity), which can quantify how tortuous an electrode is by analyzing the connection of pores within a structure. The tortuosity factor was first introduced by Epstein in 1989,[78] and can be defined for electrochemical systems by the porosity multiplied by the ratio of ionic bulk diffusion to the effective diffusion due to the tortuous path, as shown in equation 1:

$$k = \tau^2 = \epsilon \cdot \frac{D_{Bulk}}{D_{Eff}} \qquad (1)$$

where $k$ is the tortuosity factor, $\tau$ is the tortuosity, $\epsilon$ is the porosity, $D_{eff}$ is the effective diffusion coefficient, and $D_{bulk}$ a bulk diffusion coefficient[78,79]. Tortuosity in CT has gained considerable attention in the last decade[11,80,81] and is especially impactful for understanding the transport of electrolyte ions through battery electrodes. For instance, Ebner et al. used SRXTM to study the tortuosity anisotropy of three common Li-Ion electrodes with varying porosities to represent various particles shapes, (spherical, triaxial ellipsoidal, and platelet-shaped). They showed that an increased geometric tortuosity factor in the plane perpendicular to the current collection can impact the achievable LIB power density and cycling performance and predicted a factor of 4 improvement in the battery discharge rate with platelet-shaped particles in graphite electrodes[82].

## 2.2 Experimental Trends in Battery X-ray CT

Since the first battery micro-tomogram in 2001[2,35], CT has developed into a versatile technique with various CT system types offering different benefits. Laboratory-scale MicroCT is the most common CT system used in metrology and battery research, due to its large FOV (~130 $cm^2$ in the XZ direction for large scans), ease of access and use, and relatively high spatial resolution of up to ~500 $nm^{3,53,54}$. However, even with submicron resolution, MicroCT systems fall short in studying nano-scale phenomena. For this reason, battery researchers turn to SRXTM, with spatial resolutions reported in battery studies reaching as low as 50 nm[55,56]. The main compromises are the beamline time cost, increased sample preparation complexity, and limited FOV. To circumvent this, laboratory-scale NanoCT offers spatial resolutions of up to 50 nm[83], rivaling SRXTM without the necessity of high brilliance synchrotron radiation. However, there are still tradeoffs for each instrument, which should be selected carefully with the experimental goals in mind.

**Figure 2a** shows the trend in FOV, voxel size and scan duration for CT experiments in battery literature[84–97]. As many CT studies do not report the spatial resolution, the FOV was plotted versus the voxel size, since the voxel size scales with and is typically slightly less than half the spatial resolution. As shown, regardless of the battery chemistry, there is a general trend where the FOV decreases with smaller voxel sizes and is dependent on the CT system used. Indeed, in the nano-regime, synchrotron and NanoCT dominates use, while in the micro-regime, MicroCT systems are implemented more often. Additionally, in the nano-regime, NanoCT experiments are *ex-situ* due to the long scan time, and only synchrotron experiments are *in-situ* or *operando*, which illustrates the difficulty in performing dynamic experiments at a limited FOV and long scan times (**Figure 2b**).

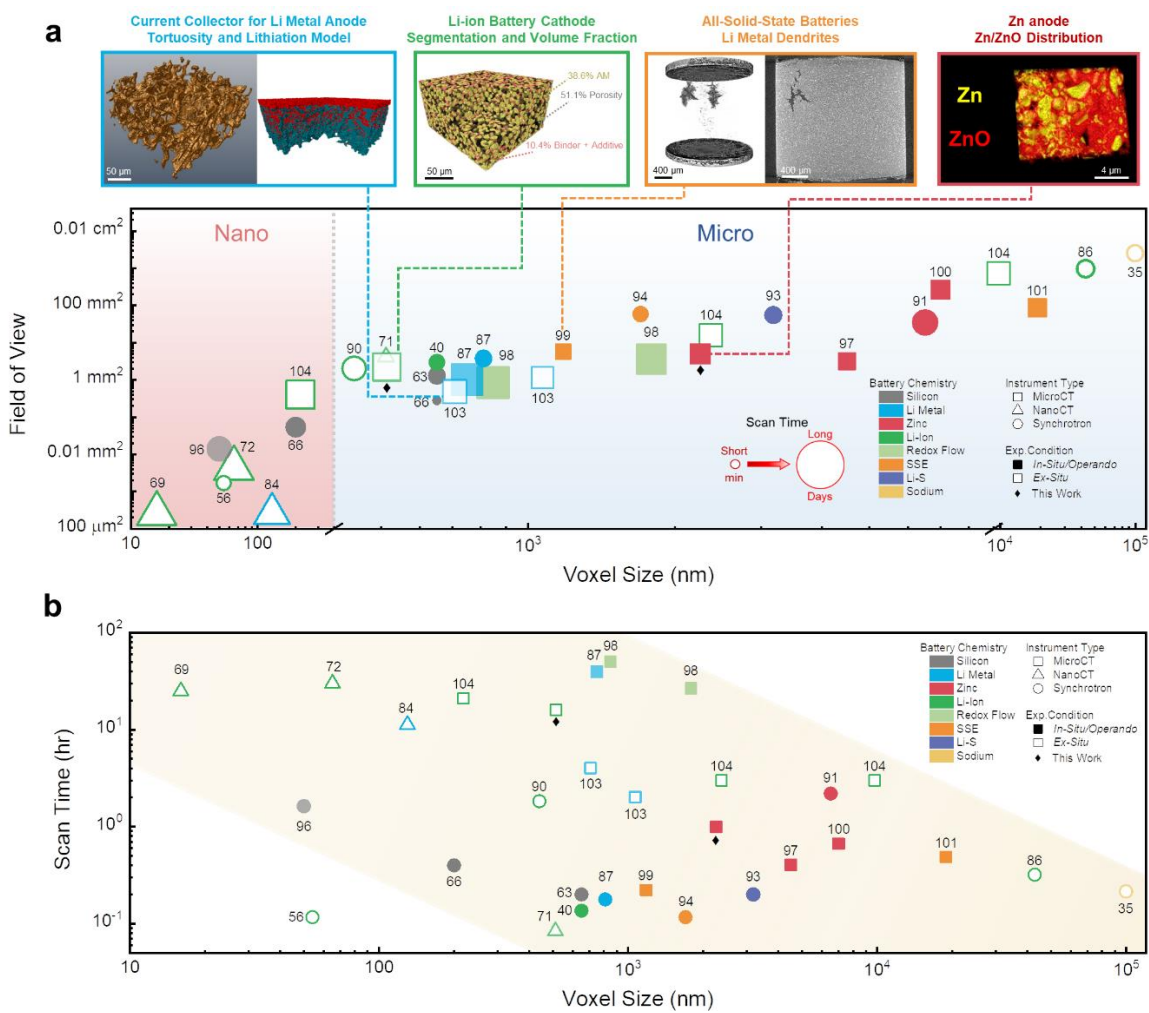

**Figure 2 | Experimental trends of computed tomography in the battery field.** Trend of a) Field of View (FOV) and b) scan time with voxel size for various battery chemistries (colors), CT systems (shape), and experimental conditions (open/filled). In a), the marker size reflects the CT scan time indicated in b).

In the micro-regime, lab-scale MicroCT dominates, and *in-situ* or *operando* experiments are quite common since such experiments tend to elucidate degradation mechanisms and key phenomena in batteries[97–101]. For many experiments, micro-sized voxels are small enough to perform novel studies for a variety of battery systems. For instance, shown in the orange in-set of **Figure 2a**, micro-sized Li-metal dendrites can be resolved in all-solid-state batteries (ASSB)[99]. Micro-sized particles on the scale of 10-100 µm can be easily resolved with MicroCT, such as the case for the Li-ion cathode and Zinc anode in the green and red insets, respectively. Particle-scale analysis can be performed for both chemistries (**Figure 3**), and *in-situ* studies are even possible as is the case for the Zinc anode[102]. Modeling can also be performed using the reconstructed scans as input. For instance, the blue inset of **Figure 2a** shows a reconstructed porous Cu current collector scanned by MicroCT and a model of the lithiation in the structure, where the porosity and tortuosity can be optimized to maximize cycle life[103].

The battery chemistry of interest also strongly influences the type of instrument used. For instance, since Li is a very light element that weakly interacts with X-rays, it is difficult to image and resolve using traditional MicroCT, which use higher X-ray energies of around 30-160 kV[3,53]. For this reason, Li-Ion battery studies shown with the green markers in **Figure 2** in the micro-regime are mainly focused on studying device structures or composite electrodes rather than investigating lithium growth. For instance, Carter *et al.* (103 in **Figure 2**) investigated the delamination in a lithium iron phosphate battery and investigated the porosity and the diffusion-based tortuosity factor of the graphite anode structure[104]. They were able to distinguish between graphite, void, and copper, but did not specifically look at Li species. Many studies looking to investigate Li with CT use ZPC with NanoCT or SRXTM systems which can use lower X-ray energies of around 5-8 keV. However, there are still constraints on the FOV and sample preparation and size for these instruments, which is why Li-ion CT studies, like Frisco *et al.* (72 in **Figure 2**)[72] and Kashkooli *et al.* (56 in **Figure 2**)[56], tend to be *ex situ*. However, custom *in-situ*/*operando* cells can be developed to help study dynamic phenomena, as shown by Vanpeene *et al.* (66 in **Figure 2**) who used X-ray CT compatible custom Swagelok cells to study volume expansion *in-situ* using SRXTM with a 200 nm voxel size[66].

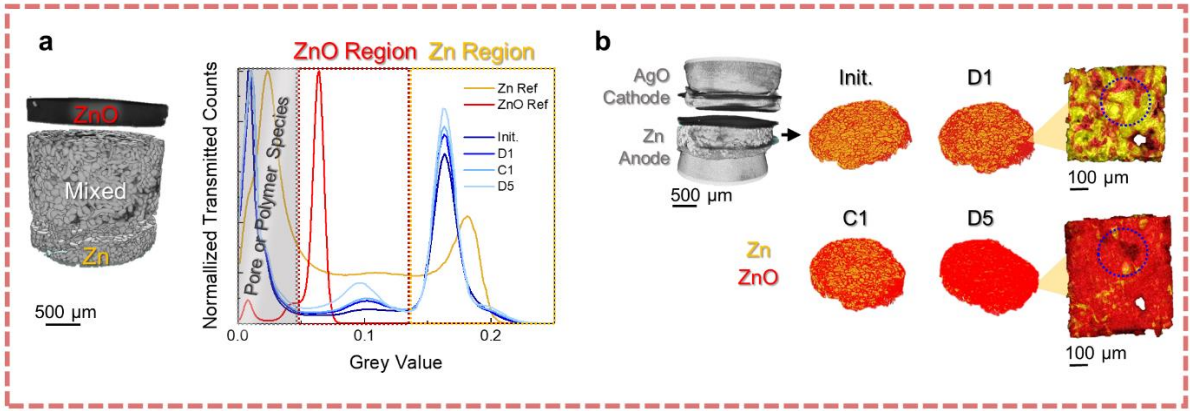

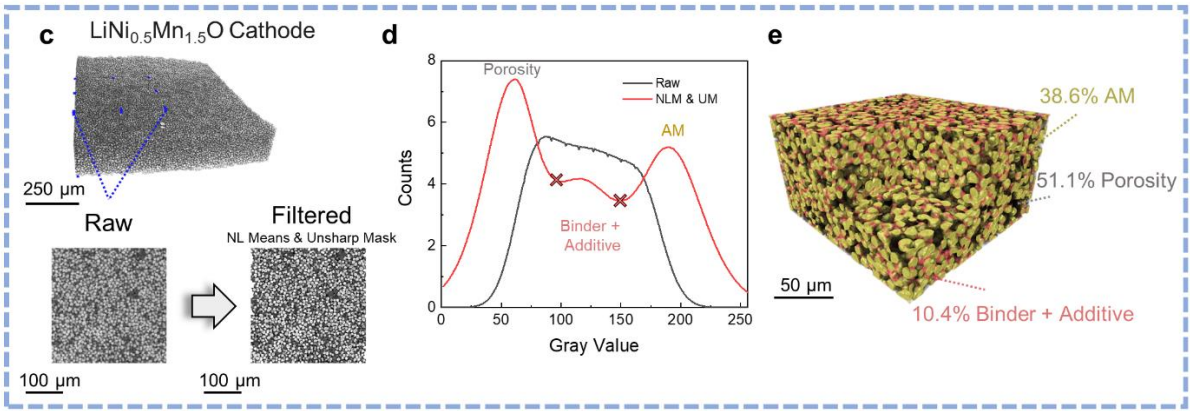

**Figure 3 | CT Segmentation and Analysis of Battery Systems.** a) Volume rendering and grey value histograms of Zn and ZnO references, and of the Zn region of the CT cell. b) Zn-AgO volume rendering with colorized 3D volume rendering of segmented Zn anode at different SOC: "Init" is initial uncycled, "D1" and "C1" is after first discharge and charge, and "D5" is after 5th cycle discharge. c) Cropped $LiNi_{0.5}Mn_{1.5}O_4$ cathode dataset and XY slices before and after Non-Local Means (NLM) and Unsharp Mask filtering (UM). d) Grey value histogram of raw data and after applying NLM and UM filters. "X" markers represent threshold regions indicated in watershed-based segmentation. e) Segmented and colorized filtered tomogram with volume fraction percentage of the 3 phases.

## 2.3 Progress and Challenges in *In-Situ* and *Operando* X-ray CT

From the onset of X-ray microtomography, it was recognized that sample preparation and design of *in-situ* environments would be an important challenge for decades to come[105]. Researchers frequently aim to achieve maximum resolution and contrast for samples that are as large as possible to achieve good statistics and obtain representative volume measurements. However, maximizing the contrast and signal-to-noise ratio for a given X-ray energy and specific composition requires limiting the sample to a specific width[106]. This is due to the attenuation of X-rays through the sample thickness, and how there is an optimal extent to which the sample of interest interacts with the incoming beam. There are many different materials used within current and next generation Li-ion batteries, but for simplicity the example of the electrode material $LiNi_{0.8}Co_{0.1}Mn_{0.1}O_2$ (NMC811) will be discussed, while attenuation coefficients for other materials can be determined from open-source databases[107]. For an X-ray energy within the range of 3 to 8 keV (typical range of lab-based NanoCT systems), the optimal thickness of a NMC811 sample is between 10 and 100 µm[106]. Synchrotron sources can tune the X-ray energies to be monochromatic within a wide range from single digit keV to energies approaching 100 keV, thus researchers can weigh the suitability of synchrotrons and beamlines for their specific application. Since most monochromatic or quasi-monochromatic sources operate in the range of 5-30 keV, the optimal width of an NMC811 electrode is between 10 µm and 1000 µm, raising the challenge of designing a small enough Li-ion cell that can achieve relevant *in-situ* or *operando* conditions. For many quantitative measurements of electrode microstructural properties, achieving a representative volume element is critical[108]. Thereafter, building an environment that facilitates *operando* or *in-situ* imaging is needed. Of most interest is electrochemical operation, but some work has focused on mechanical experiments such as *in-situ* compression of electrodes to replicate calendering[109]. Ideally, all environments would be cylindrical, achieving symmetry around the axis of rotation during imaging, which would involve circular discs of electrodes with diameters between 10-1000 µm. Specialized laser-milling has recently been shown to achieve diameters down to 80 µm with little effect on the electrode microstructure[109]. With these conditions in mind, *operando* cell environments have evolved over the past decade[110] but still suffer from design challenges that can jeopardize their performance, reliability, and operational relevance. Common issues include: damaging beam exposure[111], poor control of pressure applied on the cell, stagnant gas that causes poor ionic or electronic contact, and exposure of cell materials to contaminants including air[112]. The high impedance that is often associated with bespoke operando cell designs can limit their ability to achieve high-rate conditions necessitating modifications of well-proven cell designs like coin cells[113]. Current state-of-the-art *operando* cell designs for high-resolution imaging are based on plastic union-fittings with steel rod current collectors that seat electrodes around 1 mm in diameter[64,114,115], but much opportunity remains to improve reliability, rate performance, and ease of assembly. When a functional operational design that is suitable for the X-ray imaging conditions is achieved, further challenges await for minimizing artifacts in reconstructions, systematic errors, and data processing for quantitative analyses, such as those outlined in Boxes 1 and 2.

# 3. CT Analysis, Simulation, and Modelling

**Box 1** Artifacts and Filtering in CT

X-ray CT data analysis aims at obtaining the *truest* depiction of the sample or structure analyzed. However, experimental artifacts can distort the X-ray projections, eventually leading to data misinterpretation. As such, data containing artifacts can make segmentation challenging, thus disallowing in-depth analysis of complex structures like electrodes. Image noise is one of the most common artifacts[116], as well as cupping and streaks or dark bands from beam hardening[117–119]. As polychromatic X-rays are used in MicroCT, low energy photons are disproportionately absorbed, and the average energy of the beam increases or *hardens*. This results in cupping, where the beam is hardened more through the middle of an object than at the outer edges, and a uniformly dense material will appear non-uniform[117,119]. Streak artifacts or dark bands occur due to differences in material absorption (i.e. heavy elements next to light elements), where the beam in one area of the scan is hardened more than in another area[119]. This is especially problematic for LIBs, where light elements such as Li may be near heavier elements like Cu. Beam-hardening effects can be partly mitigated experimentally using physical filters that pre-harden the X-ray spectrum to remove low energy photons[117].

Other artifacts can originate from instrument issues or improper scan parameters. For instance, *undersampling* of the projections needed to reconstruct a sample can cause artifacts known as view and ray aliasing[119], where fine stripes appear radiating from the edges of or close to structures. View aliasing originates from a too large interval between projections and ray aliasing originates from undersampling within a given projection.

Most artifact reduction occurs during post-processing, where filtration algorithms can lessen or remove experimental artifacts and smart segmentation methods can be applied to separate out species for further analysis. Beam hardening can be treated numerically with a low-pass smoothing filter, whereby the smoothed image is used to detect large-scale intensity variations caused by beam hardening, or by a less error-prone iterative approach, which uses sequential histogram-based segmentation with grey value classification to lessen the effects of beam hardening[118,120,121].

Image noise is commonly addressed with a multitude of filtering algorithms[121]. Neighborhood statistical filters consider neighboring voxels grey values and apply a kernel operation, where voxels values are multiplied by a set of weights and then averaged over the sum of the weights to smooth or correct for noisy data[121]. Such filters are differentiated by the type of kernels used, namely mean, median, mode, minimum, maximum and gaussian filters. While filters such as these tend to blur the original data, several strategies have been developed to retain particles and pore edges, such as the Non-Local Means (NLM) and Anisotropic Diffusion (AD) filters[118,121,122], and the Unsharp Mask (UM) filter used to improve image contrast[121,123,124].

> **Box 2** CT Species Segmentation and Workflow
>
> In battery electrodes, the three main phases typically observed in an X-ray CT scan are the active material, binder, and pores. A proper segmentation of these is mandatory to ensure the quality of the extracted battery-specific parameters (such as particle size, porosity, and tortuosity). As contrast in X-ray CT is dictated by the material's X-ray absorption coefficient, the simplest segmentation method is thresholding, differentiating materials based on the numeric grey value distribution[118,125]. With Global Thresholding (GT), the segmentation is performed on the grey-value histogram of an entire 3D dataset. As can be seen in **Figure 3c-e**, filtering is a critical step, as it can reveal 3 distinct grey value regions in a Li-ion cathode (corresponding to binder, porosity, and active material), while these were undistinguishable before filtering.
>
> Manual segmentation and GT are nevertheless subject to human error and bias, and therefore a variety of automatic segmentation methods have been developed[118,125]. In contrast to GT, adaptive local segmentation methods account for neighborhood statistics to separate phases in an image. Among the multiple local segmentation methods, Bayesian Markov Random Field, Watershed, and Converging Active Contours have been shown to be the most efficient for multiclass segmentation, with tradeoffs specific to each method and sample[118].
>
> Above all, the limiting factor in segmentation is data quality, and it is crucial to have a workflow in which the dataset is optimally acquired and properly filtered to adequately define phases. The workflow can be separated into 3 stages: (1) Preprocessing (artifact removal, filtering, sharpening), (2) Segmentation (global and local thresholding), and (3) Postprocessing (denoising). Denoising algorithms are often used to prepare the dataset for structural analysis. As each stage is interconnected, filters should be chosen with the segmentation method in mind. Moreover, care must be taken as over-filtering can be an issue as well: the mean filter can introduce "unrealistic" values, and filters such as erosion, dilation, and delineation can skew multiclass data[118,126]. Therefore, knowledge of the various filters and segmentation methods is needed to ensure proper extraction of crucial morphological parameters.

### 3.1. Leveraging CT for Computational Modeling

The knowledge and large quantity of information gained from X-ray CT data leads to promising outcomes in the computational modeling of batteries. The Newman's model, a first-generation mathematical model of a lithium-ion battery, was developed in 1993[127,128]. It describes ionic transport in the concentrated electrolyte, lithium transport in the active material, and intercalation electrochemistry at the interface between active material and electrolyte. This model is supported on a 2D cartesian representation of the cell, with an extra polar coordinate dimension for the active material particles (as seen in **Figure 4** - Generation I Model) and is therefore also referred to as pseudo-2D (p2D). Since this is a 2D approach, several input parameters are necessary to consider the geometrical features of the electrodes and the cell, such as the separator thicknesses, active material particle size, the active surface area (surface area of contact between active material and electrolyte), the porosity, and tortuosity factor of both electrodes. While some of these morphological parameters (*e.g.* tortuosity factor and active surface area) are challenging to evaluate using experimental techniques[129], the stochastic generation of 3D electrode mesostructures based on the experimental parameters has been shown to be a valuable method.

To stochastically generate an electrode mesostructure, several parameters are needed: electrode composition (active material/carbon/binder volume ratio), particle radius distribution, and porosity and thickness of the electrode. In essence, the active material (typically as spheres) is generated randomly in the simulation box until the desired values are reached. Several observables, such as the amount of overlap between the spheres or the surface area, can be tuned to achieve the desired configuration. Then, the inactive phase can be added by controlling its morphology, *i.e.* as a film around the active phase or as clusters[130,131]. This approach allows access to larger electrode volumes than those experimentally achievable with NanoCT. Some commercial and academic algorithms have been reported for electrode generation[132] and for porous media analysis to extract tortuosity factors[133,134]. An alternative to the extraction of the morphological parameters is the direct use of the generated 3D electrode mesostructures in electrochemical performance models. For that purpose, the inactive phase (carbon/binder) can either be merged with the active material as the solid phase (**Figure 4** - Generation II models)[135–139] or be explicitly considered in the 3D model (**Figure 4** - Generation III models)[80,130,140,141].

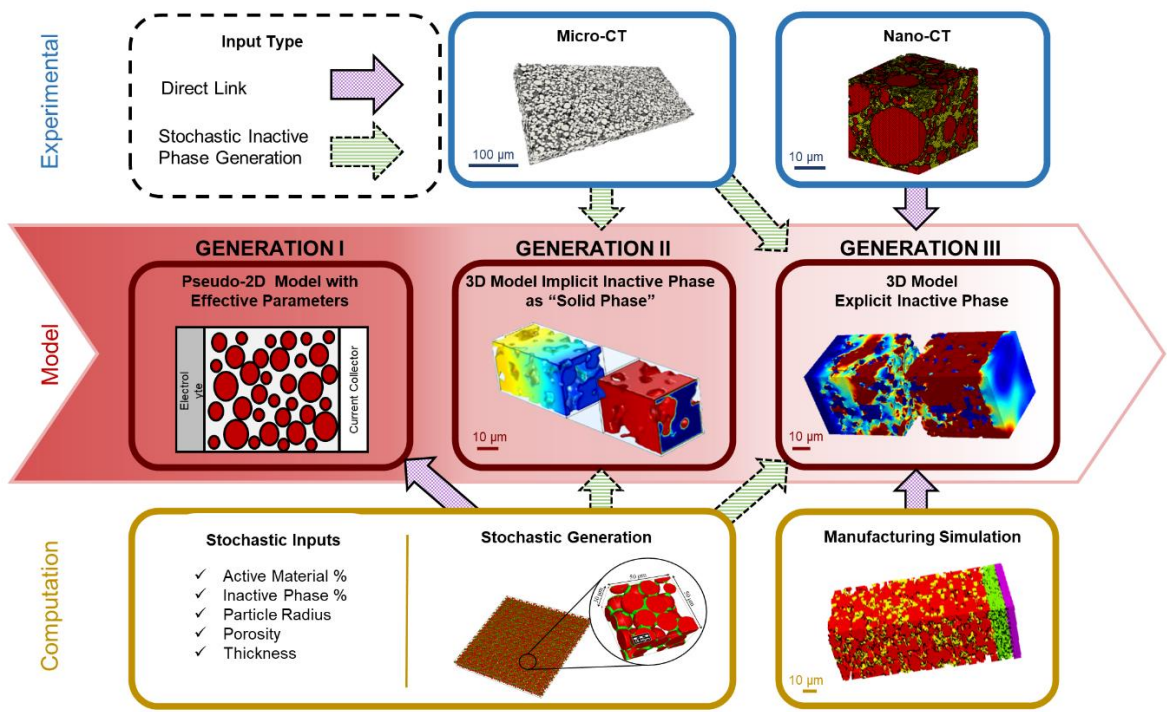

**Figure 4 | Relation between experimental tomography data, cell model, and computation of electrochemical data in battery systems.** To enable 3D models of the battery, tomography data at the micro- and nanoscale can be used either directly (purple arrows) or after stochastic inactive phase generation (green arrows). This allows to compute of the electrochemical performance of the electrode to study the effect of its morphology.

However, even full stochastic electrode mesostructure generation may not be sufficient to replace real battery electrode texture. In that sense, tomography images have been used in the recent years to increase the reliability of 3D battery cell computational models. Electrode mesostructures, reconstructed from MicroCT images, have been used for this purpose by adding the inactive phase stochastically[141,142]. Extracting the carbon and binder additives domains from MicroCT data is indeed challenging due to a too coarse spatial resolution of over ~500 nm. The dataset therefore needs further work to add the binder and carbon to the active material region. The use of MicroCT images allows the models to account for realistic active material shapes and their impact on electrochemical performance[39,143]. For instance, MicroCT has been used in redox-flow battery modeling to capture a representative volume, which is usually much larger than for lithium-ion batteries[144–146]. In this context, it has been used to predict the electrolyte impregnation and the electrochemical response of a redox flow battery using three different electrode mesostructures originating from MicroCT data[147]. However, this technique still has limitations, namely its inability to resolve the spatial location of the inactive phase in the case of lithium-ion batteries. This ability is required to investigate the impact of the arrangement of the active and inactive phases on the electrochemical, transport, and thermomechanical processes within the electrodes. For instance, the ionic transport through the electrolyte is especially impacted by the interconnectivity of pores, which requires the segmentation of the active and inactive phases to extract[75].

In NanoCT, the inactive phase can be distinguished from the active material and the porosity. The extracted structure can be directly used in the Generation III models, without any additional steps. However, the high resolution comes at the cost of a narrower FOV, resulting in a small volume of an imaged electrode. As a result, issues related to the representativeness of the volume have arisen and been addressed in the literature[148,149]. Additionally, such a multi-phase structure can be challenging to import in a finite element/volume method model, especially for numerous interfaces between a large number of phases.

Several tools in recent years have been reported to overcome this challenge[132,150,151]. In 2019, the first Generation III battery cell electrochemical model, with a positive electrode extracted from NanoCT data, was reported with an effective porosity and tortuosity for the inactive phase[152]. Furthermore, in state-of-the-art modeling, efforts have been made to limit as much as possible the use of average geometrical parameters. In 2020, several Generation III model studies have been reported, with resolved structures of the inactive phase and the ability to have no geometrical parameter as a model input[80,153]. This explicit representation of the structure is the key element to capture heterogeneities in the cell. For instance, for the 3D modeling of all solid-state batteries, locating the actual positions of voids in the electrode will be of the utmost importance to understand the device limitations.

Lastly, a new strategy is to achieve representative Generation III models without the need of tomography data: the simulation of the electrode manufacturing process[74,138,154]. By this method, the structure, from the slurry to the final calendered electrode, can be predicted with the inactive phase considered explicitly throughout the process. With the help of experimental inputs (slurry viscosity, porosity of the calendered electrode, etc...) these models are validated at each step. Despite using some geometrical approximations such as spherical particles, this approach yields satisfactory results and links experimental data with modeling, thus paving the way toward predictive digital twins of an entire manufacturing processes and showcasing its impact on battery performance predictions.

## 4. Future of Battery X-ray CT

### 4.1 Correlative Workflow Characterization

While X-ray CT is a powerful non-destructive imaging tool, it still suffers from several limitations such as the inability to distinguish chemical species with similar X-ray absorption, or to provide nano-scale information about the sample's morphology. As such, one of the main strategies to overcome these shortcomings is to combine X-ray CT with other complementary tools, i.e., correlative tomography. Several studies have already shown that both low and high-resolution X-ray CT scans can be used to determine a region of interest which is then milled using FIB/SEM. Then, volume reconstruction can be performed, and the data can be aligned with the high-resolution CT scan. Moreover, FIB/SEM benefits from the multiple detectors, such as Energy-dispersive X-ray spectroscopy (EDS), Electron Backscatter Diffraction (EBSD), Wavelength Dispersive X-ray Analysis (WDS), Raman spectroscopy, and Time-of-flight SIMS (ToF-SIMS), providing valuable insights by correlating chemical and morphological information. As already shown in the literature, a lamella of the region of interest can then be used for STEM analysis, providing nano-scale resolution imaging, combined with crystallographic and spectroscopic information thanks to Electron Energy Loss Spectroscopy (EELS) and electron diffraction. This method was successfully applied in 2014 by Burnett *et al.* to study the corrosion of stainless steel, as EBSD and EDX combined allows to determine both element segregation and grain orientation[155]. Similarly, Slater *et al.* were able to combine MicroCT with NanoCT and STEM/EDS by using Plasma FIB milling, gaining insights on the influence of grain boundaries orientation in cavity formation in Type 316 stainless steel[156].

More recently, Zubiri *et al.* demonstrated that coupling lab-scale Nano CT with electron tomography was an efficient way to combine the higher resolution of electron tomography (ET) with the wider FOV of NanoCT[157]. As such, the ML-assisted segmentation of the pores in zeolite particles from the CT data was improved significantly by using the segmentation of the higher resolution ET data as a training dataset.

In the battery field, a few studies were successful at applying correlated tomography to electrode composites or separators[11,51,158,159]. The combination of high contrast absorption X-ray tomography with ptychographic X-ray CT was shown to be able to provide detailed microstructure of Si composite anodes, distinguishing the Si particles from graphite and the carbon-binder domain, and was even able to resolve the SEI layer[51]. The obtained dataset was then used to model the state of charge distribution of individual Si particles. On the cathode side, FIB cross-sections were used to help segment the NanoCT data of a $LiNi_{0.33}Mn_{0.33}Co_{0.33}O_2$ electrode, allowing the porosity network and carbon-binder domain to be resolved[158]. This information was then successfully combined with a lower resolution MicroCT scan to evaluate the tortuosity factor of the electrode. SEM cross-section views were also combined with X-ray CT data of Li-ion battery separators to stochastically generate fibrils in the porous network. These fibrils, too small to be directly observed by MicroCT, were shown to have a significant influence on the prediction of the effective diffusion coefficient[159].

Correlative tomography, by employing a low-to-high resolution approach, is a flourishing technique that can provide the multi-scale information needed for the future of battery materials research. Combined with tools such as stochastic generation and electrochemical modeling, deep insights onto the underlying limitations of different battery systems can be gained. Nevertheless, some technical considerations still need to be addressed before this method can be widely applied to all types of systems. Principally, when working with sensitive materials, all steps of the analysis must be carried out under a protective atmosphere, necessitating careful design of the samples and transfer devices. Moreover, while the possibility to investigate a region of interest with nm-scale resolution makes this a method powerful, reducing the size of the sample sufficiently for high resolution tools (i.e., NanoCT or even TEM) can still be challenging, and tools with higher milling throughput than FIB, such as Plasma FIB (PFIB), laser PFIB, or broad ion beam milling, are often required.

**4.2 Perspective of CT Data Analysis with AI/ML**

It is evident that the future of CT data analysis is strongly correlated with the transformative tools of the emerging Digital Era, including AI/ML and multiscale modeling. ML techniques (within the wider field of AI) present a plethora of opportunities to elucidate structure-function relationships for porous electrodes images produced by CT and/or multi-physics/multi-scale modeling. In short, ML techniques give to a computer the power to learn and self-correct from data, building "models" (also called *ML models*) in an automatic way. These models can then be used to predict qualitative or quantitative outcomes, allowing for instance to unravel complex parameters interdependencies in multi-dimensional datasets and to automatize processes that would be too time-consuming to perform manually. Regarding the latter, segmenting and distinctly labelling complex features in CT-images such as cracks[12] or regions of delamination[13] can be conducted more quickly and accurately with ML techniques. The same can be said for the segmentation of composites containing multiple types of materials, such as lithium-ion battery electrodes which encompass metal oxides, carbon particles, and polymers. Such impressive segmentation capabilities allow electrochemically active, inactive materials, and pores to be distinguished faster than ever before, triggering the emergence of powerful digital twins of real electrode electrochemical operation. Still the challenge remains for CT to distinguish polymeric binders from carbon additives, both materials affecting differently the overall electrode performance. ML is also expected to help accelerate diagnostics of microstructural phenomena, as well as identification of favorable particle and electrode architectures for long life and specific operating conditions.

Applying ML Generative Adversarial Networks (GANs) to artificially create representative electrode microstructures[11] holds promise for generating 3D-resolved images with greater detail than any single imaging mode could achieve. GANs, trained with CT volumes or even slices[160], can also be used to generate *in silico* extended volumes. This would be particularly suitable for developing representative electrochemistry simulations from pre-existing tomography data of composite or electrode structures. The combination of physical-based manufacturing models and GANs can also allow the quick generation of composite structures for compositions not yet characterized, opening tremendous opportunities to accelerate the prediction and optimization of the impact of manufacturing conditions on the structures[131,161].

Critical to the acceleration of the adoption of ML techniques for these purposes is to make robust multiscale data open source, which would not only alleviate the limitation of accessing specialized imaging facilities, but also provide a wealth of microstructural information available for ML and multi-physics/multi-scale models. Such repositories should contain not only the actual data but also the metadata allowing to precisely track the conditions on which the characterization was performed, and some initiatives have already emerged[162].

We also expect the emergence of AI/ML-orchestrated workflows integrating CT characterization, data analysis, and physical model generation at multiple scales (multiscale modeling) (**Figure 5**). By coupling existing middleware technologies (*e.g.* UNICORE[163]) to AI/ML scripts, such workflows may not be difficult to develop. Thus, ML can be used to sequentially or iteratively couple different length scale models automatically with varying degrees of fidelity. Moreover, ML can also assist in comparing the modeling outcomes with experimental data. The outputs can then later be used to train AI/ML models to predict synthesis and manufacturing conditions in order to achieve optimal material properties[164]. Such an automated high-fidelity approach may revolutionize the conception of new composites materials by linking experimental data (CT) with computational simulations.

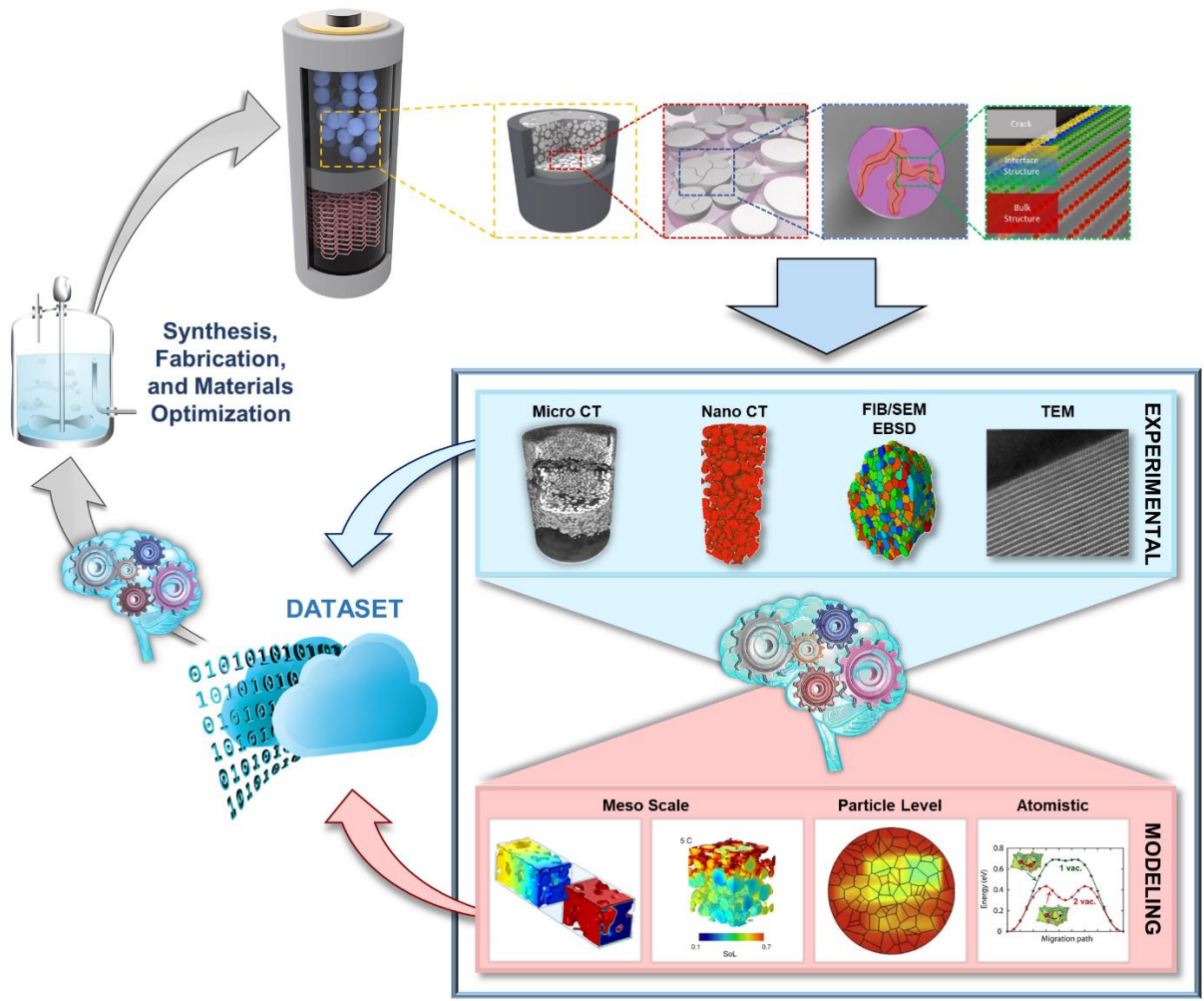

**Figure 5 | Correlative workflow analysis and modeling: combining CT and advanced characterization techniques in the development of comprehensive predicative models.** Illustrates the connection of experimental characterization data with modeling using AI/ML and the cyclical workflow to improve the synthesis, fabrication, and battery performances using predictive models.

## 5. Conclusion

Since its conception in the 1970s, CT has profoundly impacted the scientific community. In the last two decades, it has extended to greatly influence battery research and development. As a non-destructive tool, CT can perform powerful *in-situ* and *operando* studies in a multitude of battery chemistries. The reconstructed volume and extracted morphological parameters (*e.g.* particle distribution, porosity, and tortuosity) can be incorporated in predictive models to simulate battery performances. The CT images can also be used to generate large representative volumes using AI/ML techniques, such as GANs, which can generate realistic multiphase porous electrode microstructures[11]. The advent of such techniques can drastically reduce the number of required CT characterizations for 3D-resolved electrochemical models, while ensuring representative volumes for simulations.

Lastly, the combination of multiscale 3D morphological characterization techniques (*e.g.* FIB-SEM, TEM, MicroCT, and NanoCT) may pave the way for performance predictive models that can incorporate phenomena at multiple length scales. Characterization data and models can then be consolidated in open-source datasets and repositories, and even incorporated in Virtual Reality (VR) environments and tools to educate a new generation of researchers on electrode structures and associated geometric features[165]. With the tremendous progresses achieved in these last 20 years in CT experimentation, analysis, computational modeling, and AI/ML, there is promise of remarkable achievements and discoveries in the years to come.

## Acknowledgements

The authors gratefully acknowledge funding support from the US Department of Energy, Office of Basic Energy Sciences, under award number DE-SC0002357 (program manager Dr. Jane Zhu). A.A.F. and M.C. acknowledge the European Union's Horizon 2020 research and innovation program for the funding support through the European Research Council (grant agreement 772873, "ARTISTIC" project). A.A.F. acknowledges the Institut Universitaire de France for the support. This work was authored in part by the National Renewable Energy Laboratory, operated by Alliance for Sustainable Energy, LLC, for the U.S. Department of Energy (DOE) under Contract No. DEAC36-08GO28308. Funding was provided by the U.S. DOE Office of Vehicle Technology Extreme Fast Charge Program, program manager Samuel Gillard. The views expressed in the article do not necessarily represent the views of the DOE or the U.S. Government. The authors would also like to acknowledge the support of Mira Scharf in design and figure art. For the collection of the Zinc battery CT data, the authors would like to acknowledge the National Center for Microscopy and Imaging Research (NCMIR) technologies and instrumentation are supported by grant R24GM137200 from the National Institute of General Medical Sciences. AgO and Zn used in this work were provided by ZPower LLC, and $LiNi_{0.5}Mn_{1.5}O_4$ (LNMO) used in this work was supplied by Haldor Topsoe. The authors would also like to acknowledge the support of LNMO electrode fabrication by Ningbo Institute of Materials Technology and Engineering (NIMTE) in China. This work was performed in part at the San Diego Nanotechnology Infrastructure (SDNI) of UCSD, NANO3, a member of the National Nanotechnology Coordinated Infrastructure, which is supported by the National Science Foundation (Grant ECCS-1542148).

## Conflict of Interest

The authors declare no conflict of interest.

47. 2021 NSLS-II Strategic Plan. 32 (2021).

48. Chenevier, D. & Joly, A. ESRF: Inside the Extremely Brilliant Source Upgrade. *Synchrotron Radiation News* **31**, 32–35 (2018).

49. Rack, A. Hard X-ray Imaging at ESRF: Exploiting Contrast and Coherence with the New EBS Storage Ring. *Synchrotron Radiation News* **33**, 20–28 (2020).

50. Meirer, F. *et al.* Three-dimensional imaging of chemical phase transformations at the nanoscale with full-field transmission X-ray microscopy. *J Synchrotron Rad* **18**, 773–781 (2011).

51. Müller, S. *et al.* Multimodal Nanoscale Tomographic Imaging for Battery Electrodes. *Adv. Energy Mater.* **10**, 1904119 (2020).

52. Yin, L. *et al.* High Performance Printed AgO-Zn Rechargeable Battery for Flexible Electronics. *Joule* **5**, 228–248 (2021).

53. Xradia, Z. Versa 510 - Submicron X-ray Imaging: Maintain High Resolution Even at Large Working Distances. 14.

54. Varslot, T., Kingston, A., Myers, G. & Sheppard, A. High-resolution helical cone-beam micro-CT with theoretically-exact reconstruction from experimental data. *Medical Physics* **38**, 5459–5476 (2011).

55. Li, T. *et al.* Three-Dimensional Reconstruction and Analysis of All-Solid Li-Ion Battery Electrode Using Synchrotron Transmission X-ray Microscopy Tomography. *ACS Appl. Mater. Interfaces* **10**, 16927–16931 (2018).

56. Ghorbani Kashkooli, A. *et al.* Synchrotron X-ray nano computed tomography based simulation of stress evolution in LiMn2O4 electrodes. *Electrochimica Acta* **247**, 1103–1116 (2017).

57. Andrade, V. D. *et al.* Fast X-ray Nanotomography with Sub-10 nm Resolution as a Powerful Imaging Tool for Nanotechnology and Energy Storage Applications. *Advanced Materials* **n/a**, 2008653.

58. Oikonomou, C. M., Chang, Y.-W. & Jensen, G. J. A new view into prokaryotic cell biology from electron cryotomography. *Nature Reviews Microbiology* **14**, 205–220 (2016).

59. Lee, J. Z. *et al.* Cryogenic Focused Ion Beam Characterization of Lithium Metal Anodes. *ACS Energy Lett.* **4**, 489–493 (2019).

60. Seidman, D. N. Three-Dimensional Atom-Probe Tomography: Advances and Applications. *Annual Review of Materials Research* **37**, 127–158 (2007).

61. Borgia, G. C., Camaiti, M., Cerri, F., Fantazzini, P. & Piacenti, F. Study of water penetration in rock materials by Nuclear Magnetic Resonance Tomography: hydrophobic treatment effects. *Journal of Cultural Heritage* **1**, 127–132 (2000).
25